\documentclass{LMCS}

\usepackage{amssymb}
\usepackage[all]{xy}
\usepackage{listings}
\usepackage{stmaryrd}
\usepackage{changebar,xspace,hyperref,enumerate}
\mathcode`<="4268
\mathcode`>="5269
\mathchardef\ls="213C
\mathchardef\gr="213E

\newcommand{\ndi}{\mathbin{\raisebox{0.15ex}{-\hspace{-0.2em}-\hspace{-0.2em}-\hspace{-0.2em}-\hspace{-0.2em}-\hspace{-0.2em}-}\hspace{-0.7em}\succcurlyeq}}

\newcommand{\gfp}{\mbox{\rm gfp}}

\newcommand{\interpretation}[1]{\llbracket #1 \rrbracket_{\delta}}
\newcommand{\unterpretation}[1]{\llbracket #1 \rrbracket_{1}}

\newcommand{\comment}[1]{\hspace{1em}\mbox{[{\small #1}]}}

\newcommand{\bottom}{\perp}

\newcommand{\modality}{\lozenge}

\newcommand{\bigmid}{\; \bigg| \,}

\def\doi{4 (2:2) 2008}
\lmcsheading%
{\doi}
{1--23}
{}
{}
{Sep.~20, 2007}
{Apr.~09, 2008}
{}   

\begin{document}

\title[Approximating a Behavioural Pseudometric]{Approximating a
Behavioural Pseudometric without Discount for Probabilistic
Systems\rsuper *}
\author[F.~van Breugel]{Franck van Breugel\rsuper a}
\address{{\lsuper a}York University,
         4700 Keele Street, Toronto, M3J 1P3, Canada}
\email{franck@cse.yorku.ca}
\thanks{{\lsuper{a,b}}Supported by NSERC}
\author[B.~Sharma]{Babita Sharma\rsuper b}
\address{{\lsuper b}IBM Toronto Lab, 8200 Warden Avenue, Markham, L6G
1C7, Canada}
\email{babitas@ca.ibm.com}
%\thanks{{\lsuper b}Supported by NSERC}
\author[J.~Worrell]{James Worrell\rsuper c}
\address{{\lsuper c}Oxford University Computing Laboratory,
         Parks Road, Oxford, OX1 3QD, England}
\email{jbw@comlab.ox.ac.uk}
\titlecomment{{\lsuper*}An extended abstract of this paper has appeared as \cite{BSW07:fossacs}}

\begin{abstract}
\noindent
Desharnais, Gupta, Jagadeesan and Panangaden introduced a family of behavioural
pseudometrics for probabilistic transition systems.  These pseudometrics are a
quantitative analogue of probabilistic bisimilarity.  Distance zero captures
probabilistic bisimilarity.  Each pseudometric has a discount factor, a real 
number in the interval $(0, 1]$.  The smaller the discount factor, the more 
the future is discounted.  If the discount factor is one, then the future is 
not discounted at all.  Desharnais et al.\ showed that the behavioural 
distances can be calculated up to any desired degree of accuracy if the 
discount factor is smaller than one.  In this paper, we show that the 
distances can also be approximated if the future is not discounted.  A key 
ingredient of our algorithm is Tarski's decision procedure for the first order 
theory over real closed fields.  By exploiting the Kantorovich-Rubinstein 
duality theorem we can restrict to the existential fragment for which more 
efficient decision procedures exist.
\end{abstract}

\maketitle

\section{Introduction}

For systems that contain quantitative information, like, for example, 
probabilities, time and costs, several {\em behavioural pseudometrics\/} 
(and closely related notions) have been introduced (see, for example, 
\cite{BW05:tcs,B98:dc,CB02:emsoft,AHM03:icalp,DCPP05:qapl,DGJP04:tcs,DHW03:concur,GJS90:ifip,GP05a:dcec,MRST06:tcs,Y02:tcs}).  
In this paper, we focus on {\em probabilistic transition systems},
which are a variant of Markov chains.
Desharnais, Gupta, Jagadeesan and Panangaden \cite{DGJP04:tcs}
introduced a family of behavioural pseudometrics for these systems.  
These pseudometrics assign a distance, a real number in the interval $[0, 1]$,
to each pair of states of the probabilistic transition system.
The distance captures the behavioural similarity of the states.
The smaller the distance, the more alike the states behave.  The
distance is zero if and only if the states are {\em probabilistic bisimilar}, a
behavioural equivalence introduced by Larsen and Skou \cite{LS91:ic}.

The pseudometrics of Desharnais et al.\ are defined via real-valued 
interpretations of Larsen and Skou's probabilistic modal logic.  Formulae 
assume truth values in the interval $[0,1]$.  Conjunction and disjunction 
are interpreted using the lattice structure of the unit interval.  The 
modality $<a>$ is interpreted arithmetically by integration.  The behavioural 
distance between states $s_1$ and $s_2$ is then defined as the supremum over 
all formulae $\varphi$ of the difference in the truth value of $\varphi$ in 
$s_1$ and in $s_2$.%
\footnote{More generally, de Alfaro \cite{A03:concur} and McIver and
Morgan \cite{MM:tocl} 
have given real-valued interpretations to the modal mu-calculus following 
this pattern.  Moreover, de Alfaro has shown that the behavioural 
pseudometrics induced by mu-calculus formulae agree with those of 
\cite{DGJP04:tcs}.}

The definition of the behavioural pseudometrics of Desharnais et al.\
is para\-metrized by a {\em discount factor\/} $\delta$, a real number in the
interval $(0, 1]$.  The smaller the discount factor, the more (behavioural
differences in) the future are discounted.  In the case that $\delta$
equals one, the future is not discounted.  All differences in behaviour,
whether in the near or far future, contribute alike to the distance.
For systems that (in principle) run forever, we may be interested in all
these differences and, hence, in the pseudometric that does not discount
the future.

In \cite{DGJP99:concur}, Desharnais et al.\ presented an {\em algorithm\/} to
{\em approximate\/} the behavioural distances for $\delta$ smaller than one.
The first and third author \cite{BW06:tcs} presented also an
approximation algorithm for $\delta$ smaller than one.

There is a fundamental difference between pseudometrics that discount the
future and the one that does not.  This is, for example, reflected by the
fact that all pseudometrics that discount the future give rise to the same
topology, whereas the pseudometric that does not discount the future gives
rise to a different topology (see, for example, \cite[page~350]{DGJP04:tcs}).  
As a consequence,
it may not be surprising that neither approximation algorithm mentioned
in the previous paragraph can be modified in an obvious way to handle the
case that $\delta$ equals one.

The main contribution of this paper is an algorithm that approximates 
behavioural distances in case the discount factor~$\delta$ equals one. 
Starting from the \emph{logical} definition of the pseudometric by 
Desharnais et al., we first give a characterisation of the pseudometric 
as the greatest (post-)fixed point of a functional on a complete lattice 
$[0,1]^S$, where $S$ is the set of states of the probabilistic 
transition system in question.  This functional is closely related to the 
Kantorovich metric \cite{K42:dan} on probability measures.  Next, we dualize 
this characterization exploiting the Kantorovich-Rubinstein
duality theorem \cite{KR58:vestni}.  Subsequently, we show, exploiting
the dual characterization, that a pseudometric being a post-fixed
point can be expressed in the existential fragment of the first order
theory over real closed fields.  Based on the fact that this
first order theory is decidable, a result due to Tarski \cite{T51},
we show how to approximate the behavioural distances. 
Finally, we discuss an implementation of our algorithm in
Mathematica.

Exploiting the techniques put forward in this paper, we have also
developed an algorithm to approximate the behavioural pseudometric that
is presented in \cite{B05:concur}.  The other algorithm 
can be found in \cite{S06:york}.

\section{Systems and pseudometrics}

Some basic notions that will play a role in the rest of this paper are
presented below.  First we introduce the systems of interest:
probabilistic transition systems.

\begin{defi}
A {\sl probabilistic transition system\/} is a tuple $<S, \pi>$ 
consisting of
\begin{enumerate}[$\bullet$]
\item
a finite set $S$ of states and
\item
a function $\pi : S \times S \to [0, 1] \cap \mathbb{Q}$ satisfying
$\sum_{s' \in S} \pi(s, s') \in \{ 0, 1 \}$.
\end{enumerate}
We write $s \rightarrow$ if $\sum_{s' \in S} \pi(s, s') = 1$ and
$s \not\rightarrow$ if $\sum_{s' \in S} \pi(s, s') = 0$.
\end{defi}

For states~$s$ and $s'$, $\pi(s, s')$ is the probability of making a
transition to state~$s'$ given that the system is in state~$s$.
Each state~$s$ either has no outgoing transitions ($s \not\rightarrow$)
or a transition is taken with probability 1 ($s \rightarrow$).
To simplify the presentation, we do not consider the case that a 
state~$s$ may refuse to make a transition with some probability, 
that is, $\sum_{s' \in S} \pi(s, s') \in (0, 1)$.  However, all our 
results can easily be generalized to handle that case as well
(see \cite{S06:york}).  We also do not consider transitions that are 
labelled with actions.  All our results can also easily be modified to 
handle labelled transitions (see \cite{S06:york}).  In the labelled case, 
the definition of probabilistic transition system is a mild generalisation 
of the notion of Markov chain.

We restrict to rational transition probabilities in order that
probabilistic transitions systems be finitely representable.  Here we
assume that rational numbers are represented as pairs of integers in
binary. We believe that the algorithm presented in this paper could be
adapted to accommodate transition probabilities that are algebraic
numbers, but we do not pursue this question here.

In the rest of this paper, we will use the following probabilistic
transition system as our running example.

\begin{exa}
\label{example:1}
We consider a probabilistic transition system with five states:
$s_1$, $s_2$, $s_3$, $s_4$ and $s_5$.  The following table contains
the transition probabilities and, hence, captures $\pi$.
\begin{center}
\begin{tabular}{l|lllll}
      & $s_1$ & $s_2$ & $s_3$ & $s_4$ & $s_5$\\
\hline
$s_1$ & 0     & $\frac{2}{5}$ & $\frac{3}{5}$ & 0     & 0    \\
$s_2$ & $\frac{7}{10}$ & 0     & 0     & $\frac{1}{5}$ & $\frac{1}{10}$ \\
$s_3$ & 0     & 0     & 1     & 0     & 0    \\
$s_4$ & 0     & 0     & 0     & 0     & 0    \\
$s_5$ & 0     & 0     & 0     & 0     & 1    
\end{tabular}
\end{center}
The probabilistic transition system can be depicted as the following
graph.
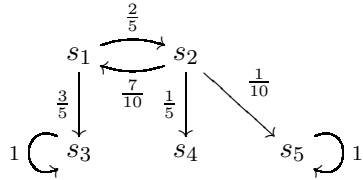
\begin{displaymath}
\UseComputerModernTips
\xymatrix{
s_1 \ar@/^/[r]^{\frac{2}{5}} \ar[d]_{\frac{3}{5}} & s_2 \ar@/^/[l]^{\frac{7}{10}} \ar[d]_{\frac{1}{5}} \ar[dr]^{\frac{1}{10}}\\
s_3 \ar@(ul,dl)_{1} & s_4 & s_5 \ar@(ur,dr)^{1}}
\end{displaymath}
\end{exa}

We consider states of a probabilistic transition system behaviourally
equivalent if they are probabilistic bisimilar \cite{LS91:ic}.

\begin{defi}
Let $<S, \pi>$ be a probabilistic transition system. An equivalence relation $\mathcal{R}$ 
on the set of states $S$ is a {\sl probabilistic bisimulation\/} if 
$s_1 \mathbin{\mathcal{R}} s_2$ implies $\sum_{s \in E} \pi(s_1, s) = \sum_{s \in E} \pi(s_2, s)$ 
for all $\mathcal{R}$-equivalence classes $E$.  States $s_1$ and $s_2$ are 
{\sl probabilistic bisimilar\/}, denoted $s_1 \sim s_2$, if $s_1 \mathbin{\mathcal{R}} s_2$ 
for some probabilistic bisimulation $\mathcal{R}$.
\end{defi}

Note that probabilistic bisimilar states $s_1$ and $s_2$ have the same probability of 
transitioning to an equivalence class $E$ of probabilistic bisimilar states.

\begin{exa}
Consider the probabilistic transition system of Example~\ref{example:1}.
The smallest equivalence relation containing $(s_3, s_5)$ is a probabilistic bisimulation.  
Hence, the states $s_3$ and $s_5$ are probabilistic bisimilar.
\end{exa}

The behavioural pseudometrics that we study in this paper yield pseudometric spaces
on the state space of probabilistic transition systems.

\begin{defi}
\label{definition:3}
A {\sl 1-bounded pseudometric space\/} is a pair $(X, d_X)$ consisting of a
set $X$ and a distance function $d_X : X \times X \to [0, 1]$ satisfying
\begin{enumerate}[(1)]
\item
for all $x \in X$, $d_X(x, x) = 0$,
\item
for all $x$, $y \in X$, $d_X(x, y) = d_X(y, x)$, and
\item
for all $x$, $y$, $z \in X$, $d_X(x, z) \leq d_X(x, y) + d_X(y, z)$.
\end{enumerate}
Instead of $(X, d_X)$ we often write $X$ and we denote the distance function
of a metric space $X$ by $d_X$.
\end{defi}

\begin{exa}
Let $X$ be a set.  The discrete metric $d_X : X \times X \to [0, 1]$
is defined by
\begin{displaymath}
d_X(x_1, x_2) = \left \{
\begin{array}{ll}
0 & \mbox{if $x_1 = x_2$}\\
1 & \mbox{otherwise.}
\end{array}
\right .
\end{displaymath}
\end{exa}

A (1-bounded) pseudometric space differs from a (1-bounded) metric space
in that different points may have distance zero in the former and not in
the latter.  Since different states of a system may behave the same, 
such states will have distance zero in our behavioural pseudometrics.

In the characterization of a behavioural pseudometric in 
Section~\ref{section:4} nonexpansive functions play a key role.

\begin{defi}
Let $X$ be a 1-bounded pseudometric space.  A function $f : X \to [0, 1]$
is {\sl nonexpansive\/} if for all $x_1$, $x_2 \in X$,
\begin{displaymath}
| f(x_1) - f(x_2) | \leq d_X(x_1, x_2).
\end{displaymath}
The set of nonexpansive functions from $X$ to $[0, 1]$ is denoted by
$X \ndi [0, 1]$.
\end{defi}

\begin{exa}
If the set $X$ is endowed with the discrete metric, then every function
from $X$ to $[0, 1]$ is nonexpansive.
\end{exa}

\section{Behavioural pseudometrics}
\label{section:3}

Desharnais, Gupta, Jagadeesan and Panangaden \cite{DGJP04:tcs} introduced 
a family of behavioural pseudometrics for probabilistic transitions systems.  
Below, we will briefly review the key ingredients of their definition.  

To define their behavioural pseudometrics, Desharnais et al.\ defined a 
real-valued semantics of a variant of Larsen and Skou's probabilistic
modal logic \cite{LS91:ic}.  We describe this variant, adapted to the case 
of unlabelled transition systems, in Definition~\ref{definition:5}.

\begin{defi}
\label{definition:5}
The logic $\mathcal{L}$ is defined by
\begin{displaymath}
\varphi ::=  \mbox{\rm true}
        \mid \modality \varphi
        \mid \varphi \wedge \varphi
        \mid \neg \varphi
        \mid \varphi \ominus q
\end{displaymath}
where $q \in [0, 1] \cap \mathbb{Q}$.
\end{defi}

The main difference between
the above logic and the one of Larsen and Skou is that we have 
$\modality \varphi$ and $\varphi \ominus q$ whereas they combine the 
operators $\modality$ and $\ominus q$ into one.  Since they consider
labelled transitions, they use the notation $<a>_q$ for this
combined operator.

Desharnais et al.\ provided a family of real-valued interpretations
of the logic.  That is, given a probabilistic transition system and
a discount factor $\delta$, the interpretation gives a quantitative
measure of the validity of a formula~$\varphi$ of the logic in a 
state~$s$ of the system.  The interpretation $\interpretation{\varphi}(s)$
is a real number in the interval $[0, 1]$.  It measures the validity
of the formula~$\varphi$ in the state~$s$.  This real number can roughly
be thought of as the probability that $\varphi$ is true in $s$.

\begin{defi}
Given a probabilistic transition system $<S, \pi>$ and 
a discount factor $\delta \in (0, 1]$, for each $\varphi \in \mathcal{L}$, 
the function $\interpretation{\varphi} : S \to [0, 1]$ is defined by
\begin{displaymath}
\begin{array}{rcl}
\interpretation{\mbox{\rm true}}(s)
& = &
1\\
\interpretation{\modality \varphi}(s)
& = &
\delta \sum_{s' \in S} \pi(s, s') \interpretation{\varphi}(s')\\
\interpretation{\varphi \wedge \psi}(s)
& = &
\min \{ \interpretation{\varphi}(s), \interpretation{\psi}(s) \}\\
\interpretation{\neg \varphi}(s)
& = &
1 - \interpretation{\varphi}(s)\\
\interpretation{\varphi \ominus q}(s)
& = &
\max \{ \interpretation{\varphi}(s) - q, 0 \}
\end{array}
\end{displaymath}
\end{defi}

\begin{exa}
Consider the probabilistic transition system of Example~\ref{example:1}.
For this system, $\interpretation{\modality \mbox{\rm true}}(s_3) = \delta$
and $\interpretation{\modality \mbox{\rm true}}(s_4) = 0$.
\end{exa}

Given a discount factor $\delta \in (0, 1]$, the behavioural pseudometric
$d_{\delta}$ assigns a distance, a real number in the interval $[0, 1]$,
to every pair of states of a probabilistic transition system.  The distance
is defined in terms of the logical formulae and their interpretation.
Roughly speaking, the distance is captured by the logical formula that
distinguishes the states the most.

\begin{defi}
Given a probabilistic transition system $<S, \pi>$ and 
a discount factor $\delta \in (0, 1]$,
the distance function $d_{\delta} : S \times S \to [0, 1]$ is defined by
\begin{displaymath}
d_{\delta}(s_{1}, s_{2})
=
\sup_{\varphi \in \mathcal{L}}
\interpretation{\varphi}(s_{1}) - \interpretation{\varphi}(s_{2}).
\end{displaymath}
\end{defi}

\begin{exa}
Consider the probabilistic transition system of Example~\ref{example:1}.
For example, the states $s_3$ and $s_4$ are $\delta$ apart.  This
distance is witnessed by the formula $\modality \mbox{\rm true}$.
The distances%
\footnote{These distances were obtained by ad-hoc methods including
  Proposition \ref{proposition:7a} and checked for numerous different
  discount factors using the algorithm described in \cite{BW06:tcs}.}
are collected in the following table.  Since a distance function is
symmetric and the distance from a state to itself is zero, we do not
give all the entries.
\begin{center}
\begin{tabular}{l|p{8em}p{8em}p{8em}p{8em}}
      & $s_1$    & $s_2$    & $s_3$    & $s_4$\\
\hline
$s_2$ & $\frac{25 \delta^2 - 2 \delta^4}{125 - 25 \delta - 35 \delta^2 + 7 \delta^3}$ \\
$s_3$ & $\frac{2 \delta^3}{25 - 7 \delta^2}$ & $\frac{5 \delta^2}{25 - 7 \delta^2}$ & \\
$s_4$ & $\delta$ & $\delta$ & $\delta$ & \\
$s_5$ & $\frac{2 \delta^3}{25 - 7 \delta^2}$ & $\frac{5 \delta^2}{25 - 7 \delta^2}$ & 0        & $\delta$
\end{tabular}
\end{center}
\end{exa}

\begin{prop}[{\cite[Theorem~5.2]{DGJP04:tcs}}]
\label{proposition:1}
$d_{\delta}$ is a 1-bounded pseudometric space.
\end{prop}
\begin{proof}
First, observe that
\begin{displaymath}
\interpretation{\varphi}(s_1) - \interpretation{\varphi}(s_2)
=
\interpretation{\neg \varphi}(s_2) - \interpretation{\neg \varphi}(s_1).
\end{displaymath}
As a consequence, we can replace 
$\interpretation{\varphi}(s_1) - \interpretation{\varphi}(s_2)$ in
the definition of $d_{\delta}$ with
$| \interpretation{\varphi}(s_1) - \interpretation{\varphi}(s_2) |$.
Checking now that $d_{\delta}$ satisfies the three conditions of
Definition~\ref{definition:3} is straightforward.
\end{proof}

States having distance zero defines an equivalence relation.  That is,
for a pseudometric $d$ on states, the relation $\equiv_d$ on states 
defined by
\begin{displaymath}
s_1 \equiv_d s_2 \mbox{ if } d(s_1, s_2) = 0
\end{displaymath}
is an equivalence relation.  We denote the equivalence class that
contains the state $s$ by $[s]_d$, that is,
\begin{displaymath}
[s]_d = \{\, s' \in S \mid d(s, s') = 0 \,\}.
\end{displaymath}

Each behavioural pseudometric $d_{\delta}$ is a quantitative analogue
of probabilistic bisimilarity.  This behavioural equivalence is
exactly captured by those states that have distance zero.

\begin{prop}[{\cite[Theorem~4.10]{DGJP04:tcs}}]
\label{proposition:2}
Given a probabilistic transition system $<S, \pi>$ and
a discount factor $\delta \in (0, 1]$, 
\begin{displaymath}
\mathord{\equiv_{d_{\delta}}} = \mathord{\sim}.
\end{displaymath}
\end{prop}
\proof%\begin{proof}
We split the proof in two parts.
\begin{enumerate}[$\bullet$]
\item
Assume that $s_1 \sim s_2$.  It suffices to show that
$\interpretation{\varphi}(s_1) = \interpretation{\varphi}(s_2)$
for all $\varphi \in \mathcal{L}$.  We can prove this by structural induction
on $\varphi$.  We focus here on the only nontrivial case: 
$\modality \varphi$.
Let $\{\, E_i \mid i \in I \,\}$ be the $\sim$-equivalence
classes.  Assume that $e_i$ is an element of $E_i$.
By induction, the function $\interpretation{\varphi}$
restricted to $E_i$ is constant.  Hence,
\begin{eqnarray*}
\interpretation{\modality \varphi}(s_1)
& = & \delta \sum_{s \in S} \pi(s_1, s) \interpretation{\varphi}(s)\\
& = & \delta \sum_{i \in I} \sum_{s \in E_i} \pi(s_1, s) \interpretation{\varphi}(s)\\
& = & \delta \sum_{i \in I} \interpretation{\varphi}(e_i) \sum_{s \in E_i} \pi(s_1, s)\\
& = & \delta \sum_{i \in I} \interpretation{\varphi}(e_i) \sum_{s \in E_i} \pi(s_2, s)
\comment{$s_1 \sim s_2$}\\
& = & \interpretation{\modality \varphi}(s_2).
\end{eqnarray*}
\item
We show that the relation $\equiv_{d_{\delta}}$ 
is a probabilistic bisimulation.  Obviously, $\equiv_{d_{\delta}}$ is
an equivalence relation.  Assume that $s_1 \mathbin{\equiv_{d_{\delta}}} s_2$.
That is, $d_{\delta}(s_1, s_2) = 0$.  Let $E$ be an $\equiv_{d_{\delta}}$-equivalence
class.  Without loss of any generality, we may assume that $E$ is of
the form $[s]_{d_{\delta}}$.  From the definition of $d_{\delta}$ we can infer that
all states in $[s]_{d_{\delta}}$ assign
the same value to each formula.  For each state $s' \not\in [s]_{d_{\delta}}$
there exists a formula $\varphi_{s'}$ such that 
$\interpretation{\varphi_{s'}}(s) \not= \interpretation{\varphi_{s'}}(s')$.
Without loss of any generality, we may assume that 
$\interpretation{\varphi_{s'}}(s) \gr \interpretation{\varphi_{s'}}(s')$.
Hence, there exists a rational $q_{s'}$ in $[0, 1]$ such that
$\interpretation{\varphi_{s'} \ominus q_{s'}}(s') = 0$ and
$\interpretation{\varphi_{s'} \ominus q_{s'}}(s) \gr 0$. 
Now consider the formula
\begin{displaymath}
\varphi = \bigwedge_{s' \not\in [s]_{d_{\delta}}} \varphi_{s'} \ominus q_{s'}.
\end{displaymath}
Then $\interpretation{\varphi}(s'') \gr 0$ iff $s'' \in [s]_{d_{\delta}}$.
As a consequence,
\begin{eqnarray*}
\lefteqn{\delta \interpretation{\varphi}(s) \sum_{s' \in [s]_{d_{\delta}}} \pi(s_1, s')}\\
& = & \delta \sum_{s' \in [s]_{d_{\delta}}} \pi(s_1, s') \interpretation{\varphi}(s')\\
& = & \delta \sum_{s'' \in S} \pi(s_1, s'') \interpretation{\varphi}(s'')
\comment{$\interpretation{\varphi}(s'') = 0$ for all $s'' \not\in [s]_{d_{\delta}}$}\\
& = & \interpretation{\modality \varphi}(s_1)\\
& = & \interpretation{\modality \varphi}(s_2) 
\comment{$d_{\delta}(s_1, s_2) = 0$}\\
& = & \delta \interpretation{\varphi}(s) \sum_{s' \in [s]_{d_{\delta}}} \pi(s_2, s').
\end{eqnarray*}
Therefore, $\sum_{s' \in [s]_{d_{\delta}}} \pi(s_1, s') = \sum_{s' \in [s]_{d_{\delta}}} \pi(s_2, s')$
and, hence, $\equiv_{d_{\delta}}$ is a probabilistic bisimulation.\qed
\end{enumerate}

In \cite{DGJP99:concur}, Desharnais et al.\ present a decision procedure
for the behavioural pseudometric $d_{\delta}$ when $\delta$ is 
smaller than one.  Let us briefly sketch their algorithm.  They
define the depth of a logical formula as follows.
\begin{displaymath}
\begin{array}{rcl}
\mbox{\rm depth}(\mbox{\rm true}) & = & 0\\
\mbox{\rm depth}(\modality \varphi) & = & \mbox{\rm depth}(\varphi) + 1\\
\mbox{\rm depth}(\varphi \wedge \psi) & = & \max \{ \mbox{\rm depth}(\varphi), \mbox{\rm depth}(\psi) \}\\
\mbox{\rm depth}(\neg \varphi) & = & \mbox{\rm depth}(\varphi)\\
\mbox{\rm depth}(\varphi \ominus q) & = & \mbox{\rm depth}(\varphi)
\end{array}
\end{displaymath}
One can easily verify that 
$\interpretation{\varphi}(s_{1}) - \interpretation{\varphi}(s_{2}) \leq \delta^{\mbox{\rm \small depth}(\varphi)}$
for each $\varphi \in \mathcal{L}$.
This suggests that one can compute $d_{\delta}$ to any desired degree of 
accuracy by restricting attention to formulae $\varphi$ of a fixed modal 
depth. Clearly, there exist infinitely many formulae of each fixed modal 
depth.  Nevertheless, Desharnais et al.\ show how to construct a finite 
subset $\mathcal{F}_n$ of the logical formulae of at most depth $n$ such 
that
\begin{displaymath}
d_{\delta}(s_1, s_2) - \sup_{\varphi \in \mathcal{F}_n} \interpretation{\varphi}(s_{1}) - \interpretation{\varphi}(s_{2}) \leq \delta^n.
\end{displaymath}
In this way, $d_{\delta}(s_1, s_2)$ can be approximated up to arbitrary accuracy
\emph{provided} $\delta$ is smaller than one.

\section{A fixed point characterization and its dual}
\label{section:4}

For the rest of this paper, we focus on the behavioural pseudometric
that does not discount the future.  That is, we concentrate on the
pseudometric~$d_1$.  Below, we present an alternative characterization
of this pseudometric.  In particular, we characterize $d_1$ as the
greatest (post-)fixed point of a function $\Delta$ from a complete
lattice to itself.  This characterization can be viewed as a
quantitative analogue of the greatest fixed point characterization of
bisimilarity \cite{P81:tcs}.

We also dualize the definition of $\Delta$ exploiting the Kantorovich-Rubinstein
duality theorem \cite{KR58:vestni}.  As we will see in Section~\ref{section:6},
this dual characterization will allow us to define $\Delta$ as the solution to a minimization
problem rather than a maximization problem, as above.  In turn this will allow us to capture
the fact that a pseudometric is a post-fixed point of $\Delta$ in the existential fragment
of the first order theory over real closed fields.

For the rest of this paper, we fix a probabilistic transition
system $<S, \pi>$.  We endow the set of pseudometrics on $S$ with
the following order.

\begin{defi}
The relation $\sqsubseteq$ on 1-bounded pseudometrics on $S$ is defined by
\begin{displaymath}
d_1 \sqsubseteq d_2 \mbox{ if } d_1(s_1, s_2) \geq d_2(s_1, s_2)
\mbox{ for all $s_1$, $s_2 \in S$}.
\end{displaymath}
\end{defi}

Note the reverse direction of $\sqsubseteq$ and $\geq$ in the above
definition.  We decided to make this reversal so that $d_1$ is a
greatest fixed point, in analogy with the characterization of bisimilarity,
rather than a least fixed point.  This choice has no impact on any
results in this paper.

\begin{prop}[{\cite[Lemma~3.2]{DGJP02:lics}}]
\label{proposition:3}
The set of 1-bounded pseudometrics on $S$ endowed with the order
$\sqsubseteq$ forms a complete lattice.
\end{prop}
\begin{proof}
Obviously, $\sqsubseteq$ is a partial order.  The top element is
the 1-bounded pseudometric $\top$ defined by
\begin{displaymath}
\top(s_1, s_2) = 0.
\end{displaymath}
The bottom element is the 1-bounded pseudometric $\bottom$ defined by
\begin{displaymath}
\bottom(s_1, s_2) = \left \{
\begin{array}{ll}
0 & \mbox{if $s_1 = s_2$}\\
1 & \mbox{otherwise.}
\end{array}
\right .
\end{displaymath}
Let $D$ be a nonempty set of 1-bounded pseudometrics on $S$.  The
meet of $D$ is the 1-bounded pseudometric $\bigsqcap D$ defined by
\begin{displaymath}
(\bigsqcap D)(s_1, s_2)
=
\sup_{d \in D} d(s_1, s_2).
\end{displaymath}
The join of $D$ can be expressed in terms of the meet of $D$
(see, for example, \cite[Lemma~2.15]{DP90}).
\end{proof}

Whereas meets of pseudometrics are computed pointwise using the supremum
on [0,1], joins of pseudometrics are not.

Next, we introduce a function from this complete lattice to itself
of which the behavioural pseudometric $d_1$ is the greatest fixed point.

\begin{defi}
\label{definition:9}
Let $d$ be a 1-bounded pseudometric on $S$.  The
distance function $\Delta(d) : S \times S \to [0, 1]$ is defined by
\begin{displaymath}
\Delta(d)(s_1, s_2)
=
\max\, \left \{\, \sum_{s \in S} f(s) (\pi(s_1, s) - \pi(s_2, s)) \bigmid f \in (S, d) \ndi [0, 1] \, \right \}
\end{displaymath}
if $s_1 \rightarrow$ and $s_2 \rightarrow$, and
$
\Delta(d)(s_1, s_2) = \left \{
\begin{array}{ll}
0
& \mbox{if $s_1 \not\rightarrow$ and $s_2 \not\rightarrow$}\\
1
& \mbox{otherwise.}
\end{array}
\right .  $ 
\end{defi}

Note that we can write $\max$ above rather than $\sup$
since $(S, d) \ndi [0, 1]$, being a closed subset of the product space
$[0,1]^S$, is compact.

The functional $\Delta$ is closely related to the Kantorovich metric 
\cite{K42:dan} on probability measures.  In the definition of that
metric, nonexpansive functions play a key role.%
\footnote{The Kantorovich metric is the smallest distance function
on probability measures for which integration of nonexpansive
functions is nonexpansive.}

\begin{prop}
$\Delta(d)$ is a 1-bounded pseudometric on $S$.
\end{prop}
\begin{proof}
Note that $f \in (S, d) \ndi [0, 1]$ implies $1 - f \in (S, d) \ndi [0, 1]$.
Furthermore, if $s_1 \rightarrow$ and $s_2 \rightarrow$ then
\begin{eqnarray*}
\lefteqn{\sum_{s \in S} (1 - f)(s)(\pi(s_1, s) - \pi(s_2, s))}\\
& = & \sum_{s \in S} \pi(s_1, s) - \sum_{s \in S} \pi(s_2, s) + \sum_{s \in S} f(s) (\pi(s_2, s) - \pi(s_1, s))\\
& = & \sum_{s \in S} f(s) (\pi(s_2, s) - \pi(s_1, s))\\
& = & \sum_{s \in S} f(s) \pi(s_2, s) - \sum_{s \in S} f(s) \pi(s_1, s).
\end{eqnarray*}
As a consequence, if $s_1 \rightarrow$ and $s_2 \rightarrow$ then
\begin{displaymath}
\Delta(d)(s_1, s_2) =
\max\, \left \{\, \left | \sum_{s \in S} f(s) \pi(s_1, s) - \sum_{s \in S} f(s) \pi(s_2, s) \right | \bigmid f \in (S, d) \ndi [0, 1] \, \right \}.
\end{displaymath}
Now that we have this alternative representation of $\Delta(d)$, 
checking that it satisfies the three conditions of Definition~\ref{definition:3}
is straightforward.
\end{proof}

\begin{prop}[{\cite[Proposition~38]{BHMW06:tcs}}]
\label{proposition:5}
$\Delta$ is order-preserving.
\end{prop}
\begin{proof}
Let $d_1$ and $d_2$ be 1-bounded pseudometrics on $S$ with $d_1 \sqsubseteq d_2$.
Note that any function $S \rightarrow [0,1]$ that is nonexpansive with
respect to $d_2$ is also nonexpansive with respect to $d_1$.  Therefore
$\Delta(d_2)(s_1,s_2) \leq \Delta(d_1)(s_1,s_2)$ for all $s_1,s_2 \in
S$ since the latter involves taking the $\max$ over a larger set.
%Let $d_1$ and $d_2$ be 1-bounded pseudometrics on $S$ with $d_1 \sqsubseteq d_2$.  
%
%Assume that $f \in (S, d_2) \ndi [0, 1]$.
%Then 
%\begin{eqnarray*}
%\lefteqn{| f(s_1) - f(s_2) |}\\
%& \leq & d_2(s_1, s_2)
%\comment{$f \in (S, d_2) \ndi [0, 1]$}\\
%& \leq & d_1(s_1, s_2)
%\comment{$d_1 \sqsubseteq d_2$}
%\end{eqnarray*}
%As a consequence,
%\begin{equation}
%\label{equation:1}
%(S, d_1) \ndi [0, 1] \supseteq (S, d_2) \ndi [0, 1].
%\end{equation}
%
%We have to show that 
%$\Delta(d_1)(s_1, s_2) \geq \Delta(d_2)(s_1, s_2)$.  We focus on the only
%nontrivial case: $s_1 \rightarrow$ and $s_2 \rightarrow$.
%In this case,
%\begin{eqnarray*}
%\lefteqn{\Delta(d_1)(s_1, s_2)}\\
%& = & \max\, \left \{\, \sum_{s \in S} f(s) (\pi(s_1, s) - \pi(s_2, s)) \bigmid f \in (S, d_1) \ndi [0, 1] \, \right \}\\
%& \geq & \max\, \left \{\, \sum_{s \in S} f(s) (\pi(s_1, s) - \pi(s_2, s)) \bigmid f \in (S, d_2) \ndi [0, 1] \, \right \}
%\comment{(\ref{equation:1})}\\
%& = & \Delta(d_2)(s_1, s_2).
%\end{eqnarray*}
\end{proof}

Since $\Delta(d)$ is a 1-bounded pseudometric on $S$ and $\Delta$ is 
order-preserving, we can conclude from Tarski's fixed point theorem 
\cite[Theorem~1]{T55:pjm} that $\Delta$ has a greatest fixed point.
We denote the greatest fixed point of $\Delta$ by $\gfp(\Delta)$.
This greatest fixed point of $\Delta$ is also the greatest post-fixed
point of $\Delta$ (see, for example, \cite[Theorem~4.11]{DP90}%
\footnote{$d$ is a {\em post-fixed point\/} of $\Delta$ if $d \sqsubseteq \Delta(d)$.
In \cite[page~94]{DP90}, such a $d$ is called a pre-fixpoint.}).

\begin{thm}
\label{proposition:6}
$d_1 = \gfp(\Delta)$.
\end{thm}

\begin{proof}
We first prove that $d_1$ is a post-fixed point of $\Delta$.  
That is, we show that $\Delta(d_1)(s_1, s_2) \leq d_1(s_1, s_2)$.  
To prove this, we distinguish the following three cases.
\begin{enumerate}[$\bullet$]
\item
If $s_1 \not\rightarrow$ and $s_2 \not\rightarrow$ then 
the property is vacuously true.
\item
If $s_1 \not\rightarrow$ and $s_2 \rightarrow$, or
$s_1 \rightarrow$ and $s_2 \not\rightarrow$, then
the formula $\modality \mbox{\rm true}$ witnesses that the
states $s_1$ and $s_2$ have distance one.
\item
Assume that $s_1 \rightarrow$ and $s_2 \rightarrow$.
According to \cite[Proposition~39]{BW05:tcs}, the set
$\{\, \unterpretation{\varphi} \mid \varphi \in \mathcal{L} \,\}$
is dense in $(S, d_1) \ndi [0, 1]$, that is,
each $f \in (S, d_1) \ndi [0, 1]$ can be approximated up to
arbitrary accuracy by some $\unterpretation{\varphi}$.  As a consequence,
\begin{eqnarray*}
\lefteqn{\max\, \left \{\, \sum_{s \in S} f(s) (\pi(s_1, s) - \pi(s_2, s)) \bigmid f \in (S, d_1) \ndi [0, 1] \, \right \}}\\
& = & \max\, \left \{\, \sum_{s \in S} \unterpretation{\varphi}(s) (\pi(s_1, s) - \pi(s_2, s)) \bigmid \varphi \in \mathcal{L} \, \right \}\\
& = & \max\, \left \{\, \sum_{s \in S} \pi(s_1, s) \unterpretation{\varphi}(s) - \sum_{s \in S} \pi(s_2, s) \unterpretation{\varphi}(s) \bigmid \varphi \in \mathcal{L} \, \right \}\\
& = & \max\, \left \{\, \unterpretation{\modality \varphi}(s_1) - \unterpretation{\modality \varphi}(s_2) \bigmid \varphi \in \mathcal{L} \, \right \}\\
& \leq & d_1(s_1, s_2).
\end{eqnarray*}
\end{enumerate}

Next we prove that $d_1$ is the greatest post-fixed point of $\Delta$.
Assume that $d$ is a post-fixed point of $\Delta$.  We have to show that
$d \sqsubseteq d_1$.  That is, $d_1(s_1, s_2) \leq d(s_1, s_2)$. 
We restrict our attention to the case that
$s_1 \rightarrow$ and $s_2 \rightarrow$.  It suffices to show that
\begin{displaymath}
\unterpretation{\varphi}(s_1) - \unterpretation{\varphi}(s_2) \leq d(s_1, s_2)
\end{displaymath}
for all $\varphi \in \mathcal{L}$.  This can be proved by structural induction on
$\varphi$.  We consider only the nontrivial case: $\modality \varphi$.
\begin{eqnarray*}
\lefteqn{\unterpretation{\modality \varphi}(s_1) - \unterpretation{\modality \varphi}(s_2)}\\
& = & \sum_{s \in S} \pi(s_1, s) \unterpretation{\varphi}(s) - \sum_{s \in S} \pi(s_2, s) \unterpretation{\varphi}(s)\\
& = & \sum_{s \in S} \unterpretation{\varphi}(s) (\pi(s_1, s) - \pi(s_2, s))\\
& \leq & \max \left \{\, \sum_{s \in S} f(s) (\pi(s_1, s) - \pi(s_2, s)) \bigmid f \in (S, d) \ndi [0, 1] \, \right \}\\
&& \comment{by induction, $\unterpretation{\varphi} \in (S, d) \ndi [0, 1]$}\\
& = & \Delta(d)(s_1, s_2)\\
& \leq & d(s_1, s_2)
\comment{$d$ is a post-fixed point of $\Delta$}
\end{eqnarray*}
\end{proof}

A similar result can be obtained by combining Theorem~40 and 44 of \cite{BHMW06:tcs}.

Let us recall (a minor variation of) the Kantorovich-Rubinstein duality theorem.  Let $X$ be a 
1-bounded compact pseudometric space.  Let $\mu_1$ and $\mu_2$ be Borel probability measures on $X$.  
We denote the set of Borel probability measures on the product space with marginals 
$\mu_1$ and $\mu_2$, that is, the Borel probability measures $\mu$ on $X^2$ such that 
for all Borel subsets $B$ of $X$,
\begin{displaymath}
\mu(B \times X) = \mu_1(B) \mbox{ and } \mu(X \times B) = \mu_2(B),
\end{displaymath}
by $\mu_1 \otimes \mu_2$.  The Kantorovich-Rubinstein duality theorem tells us
\begin{displaymath}
\max \left \{ \int_X f d\mu_1 - \int_X f d\mu_2 \bigmid f \in X \ndi [0, 1]  \right\}
=
\min \left \{ \int_{X^2} d_X d\mu \bigmid \mu \in \mu_1 \otimes \mu_2  \right\}.
\end{displaymath}

The following proposition, which is a consequence of the Kantorovich-Rubinstein 
duality theorem, defines $\Delta(d)$ as a minimum as opposed to the maximum in 
Definition~\ref{definition:9}.

\begin{prop}[{\cite[Corollary~19]{BW06:tcs}}]
\label{proposition:7}
Let $d$ be a 1-bounded pseudometric on $S$.  Let $s_1$, $s_2 \in S$
such that $s_1 \rightarrow$ and $s_2 \rightarrow$.  Then
\begin{displaymath}
\Delta(d)(s_1, s_2) 
=
\min\, \left \{\, \sum_{(s_i, s_j) \in S^2} d(s_i, s_j) \mu(s_i, s_j) \bigmid
\mu \in \pi(s_1, \cdot) \otimes \pi(s_2, \cdot) \, \right \}
\end{displaymath}
where $\mu \in \pi(s_1, \cdot) \otimes \pi(s_2, \cdot)$ if
\begin{displaymath}
\forall s_j \in S \sum_{s_i \in S} \mu(s_i, s_j) = \pi(s_1, s_j) \wedge
\forall s_i \in S \sum_{s_j \in S} \mu(s_i, s_j) = \pi(s_2, s_i).
\end{displaymath}
\end{prop}
\begin{proof}
Since the set $S$ is finite, the space $(S, d)$ is compact.  The probability
distributions $\pi(s_1, \cdot)$ and $\pi(s_2, \cdot)$ define Borel
probability measures on $(S, d)$.  Applying the Kantorovich-Rubinstein
gives us the desired result.
\end{proof}

\section{The algorithm}
\label{section:6}

Before we present our algorithm, we first show that the fact that a pseudometric
is a post-fixed point of $\Delta$ can be expressed in (the existential 
fragment of) the first order theory over real closed fields.
This will allow us to exploit Tarski's decision procedure to
approximate the behavioural pseudometric.  

For the rest of this paper, we assume that the probabilistic transition system $<S, \pi>$
has $N$ states $s_1$, $s_2$, \ldots, $s_N$.  Instead of $\pi(s_i, s_j)$ 
we will write $\pi_{ij}$.  We represent a 1-bounded
pseudometric on the set $S$ of states of the probabilistic transition
system, as (the values of) a collection of real valued variables $d_{ij}$. 

The fact that $d$ is a 1-bounded pseudometric can now be captured as follows.

\begin{defi}
The predicate $\mbox{\rm pseudo}(d)$ is defined by
\begin{eqnarray*}
\mbox{\rm pseudo}(d) 
& \equiv & \bigwedge_{1 \le i, j \le N} d_{ij} \ge 0 \wedge d_{ij} \le 1 \wedge\\
&        & \bigwedge_{1 \le i \le N} d_{ii}  = 0 \wedge
           \bigwedge_{1 \le i, j \le N} d_{ij}  = d_{ji} \wedge
           \bigwedge_{1 \le h, i, j \le N} d_{hj} \le d_{hi} + d_{ij}
\end{eqnarray*}
\end{defi}

Furthermore, the fact that $d$ is a post-fixed point of $\Delta$ can be captured as follows.

\begin{defi}
The predicate $\mbox{\rm post-fixed}(d)$ is defined by
\begin{eqnarray*}
\lefteqn{\mbox{\rm post-fixed}(d)}\\
& \equiv &
\bigwedge_{1 \le i_0, j_0 \le N}
\mbox{\rm post-fixed}_1(d, i_0, j_0) \vee
\mbox{\rm post-fixed}_2(d, i_0, j_0) \vee
\mbox{\rm post-fixed}_3(d, i_0, j_0)
\end{eqnarray*}
where
\begin{eqnarray*}
\mbox{\rm post-fixed}_1(d, i_0, j_0)
& \equiv & \sum_{1 \le i \le N} \pi_{i_0 i} \gr 0 \wedge
           \sum_{1 \le j \le N} \pi_{j_0 j} \gr 0 \wedge\\
&        & \exists (\mu_{ij})_{1\leq i,j \leq N} \bigwedge_{1 \leq i, j \leq N} \mu_{ij} \geq 0 \wedge \mu_{ij} \leq 1\\
&        & \phantom{\exists (\mu_{ij})_{1\leq i,j \leq N}} \bigwedge_{1 \le j \le N} \sum_{1 \le i \le N} \mu_{ij} = \pi_{i_0 j} \wedge\\
&        & \phantom{\exists (\mu_{ij})_{1\leq i,j \leq N}} \bigwedge_{1 \le i \le N} \sum_{1 \le j \le N} \mu_{ij} = \pi_{j_0 i} \wedge\\
&        & \phantom{\exists (\mu_{ij})_{1\leq i,j \leq N}} \sum_{1 \le i, j \le N} d_{ij} \mu_{ij} \le d_{i_0 j_0}\\
\mbox{\rm post-fixed}_2(d, i_0, j_0)
& \equiv & \sum_{1 \le i \le N} \pi_{i_0 i} = 0 \wedge
           \sum_{1 \le j \le N} \pi_{j_0 j} = 0 \wedge
           0 \le d_{i_0 j_0}\\
\mbox{\rm post-fixed}_3(d, i_0, j_0)
& \equiv & \left ( \left (\sum_{1 \le i \le N} \pi_{i_0 i} \gr 0 \wedge
             \sum_{1 \le j \le N} \pi_{j_0 j} = 0 \right ) \vee \right . \\
&& \phantom{\left ( \right .} \left .
            \left (\sum_{1 \le i \le N} \pi_{i_0 i} = 0 \wedge
             \sum_{1 \le j \le N} \pi_{j_0 j} \gr 0 \right ) \right ) \wedge\\
&&         1 \le d_{i_0 j_0}
\end{eqnarray*}
\end{defi}

Now we are ready to present our algorithm.  Consider the states
$s_{i_0}$ and $s_{j_0}$.  We restrict our attention to the case 
that $s_{i_0} \rightarrow$ and $s_{j_0} \rightarrow$.  In the
other cases the computation of the distance is trivial.

In our algorithm, we use the algorithm {\tt tarski} that takes 
as input a sentence of the first order theory of real closed fields 
and decides the truth or falsity of the given sentence.  The fact 
that there exists such an algorithm was first proved by Tarski \cite{T51}.

Let $\epsilon$ be the desired accuracy.  That is, we want to find an 
interval $[\ell_0, u_0] \subseteq [0, 1]$ such that $u_0 - \ell_0 \le \epsilon$ 
and $d_1(s_{i_0}, s_{j_0}) \in [\ell_0, u_0]$.
The algorithm {\tt approximate} takes as input an interval
$[\ell, u] \subseteq [0, 1]$ such that $d_1(s_{i_0}, s_{j_0}) \in [\ell, u]$
and returns the desired result.  As a consequence, {\tt approximate}(0, 1)
returns an approximation of $d_1(s_{i_0}, s_{j_0})$ with accuracy $\epsilon$.

\begin{quote}
\lstset{basicstyle=\ttfamily,basewidth={0.5em,0.1em},breaklines=true,mathescape=true}
\begin{lstlisting}
approximate($\ell$, $u$):
   if $u - \ell \le \epsilon$
      return $[\ell, u]$
   else
      $m = \frac{\ell + u}{2}$
      if tarski($\exists d\, \mbox{\rm pseudo}(d) \wedge \mbox{\rm post-fixed}(d) \wedge d_{i_0 j_0} \le m$)
         return approximate($\ell$, $m$)
      else
         return approximate($m$, $u$)
\end{lstlisting}
\end{quote}

Note that the argument of {\tt tarski} is a sentence that is part of
the existential fragment of the first order theory over real closed fields.
For this fragment there are more efficient decision procedures than
for the general theory (see, for example, \cite{SPR96:jacm}).  

Let us sketch a correctness proof of our algorithm.  Assume that
$d_1(s_{i_0}, s_{j_0}) \in [\ell, u]$.  We distinguish the following
three cases.
\begin{enumerate}[$\bullet$]
\item
If $u - \ell \le \epsilon$, then the algorithm obviously returns the
desired result.
\item
Assume that $u - \ell \gr \epsilon$ and suppose that
{\tt tarski}
returns true.  Then there exists a 1-bounded pseudometric $d$ that is 
a post-fixed point of $\Delta$ and $d(s_{i_0}, s_{j_0}) \le m$.  Since $d_1$
is the greatest post-fixed point of $\Delta$, we have that $d \sqsubseteq d_1$.
Hence, $d_1(s_{i_0}, s_{j_0}) \leq d(s_{i_0}, s_{j_0}) \le m$.  By assumption
$d_1(s_{i_0}, s_{j_0}) \in [\ell, u]$, therefore $d_1(s_{i_0}, s_{j_0}) \in [\ell, m]$.
\item
Assume that $u - \ell \gr \epsilon$ and suppose that {\tt tarski}
returns false.  Then $d(s_{i_0}, s_{j_0}) \gr m$ for every 1-bounded
pseudometric $d$ that is a post-fixed point of $\Delta$.  Since $d_1$
is a post-fixed point of $\Delta$, we have that $d_1(s_{i_0}, s_{j_0})
\gr m$.  By assumption $d_1(s_{i_0}, s_{j_0}) \in [\ell, u]$,
therefore, $d_1(s_{i_0}, s_{j_0}) \in [m, u]$.
\end{enumerate}
Obviously, the algorithm terminates.

\section{Conclusion}

This paper combines a number of ingredients, known already for a long time, 
including the Kantorovich-Rubinstein duality theorem of the fifties, 
Tarski's fixed point theorem of the forties and Tarski's decision procedure 
for the first order theory of real closed fields of the thirties.
We show that the behavioural pseudometric $d_1$, which does not discount 
the future, can be approximated up to an arbitrary accuracy.  While the 
combination of the above results into a decision procedure for the 
pseudometric is not technically difficult, we do solve a problem that has 
been open since 1999.  Most of the results in Section~\ref{section:3} and 
\ref{section:4} are (variations on) known results.  As far as we know, 
%Proposition~\ref{proposition:6a} and 
the results in Section~\ref{section:6} and Appendix~\ref{section:8} are new.
The techniques exploited in this paper have also been used to
approximate other behavioural pseudometrics that do not discount the
future such as, for example, the one presented in \cite{B05:concur}.
Furthermore, our algorithm can easily be adjusted to the discounted
case.  

Since the satisfiability
problem for the existential fragment of the first order theory of
real closed fields is in \textsc{PSPACE}, it is not surprising that
our algorithm can only handle small examples as we have shown in
Appendix~\ref{section:8}.  As a consequence, the quest for practical
algorithms to approximate $d_1$ is still open.  Since the closure
ordinal of $\Delta$ is $\omega$, as proved in
Appendix~\ref{section:added}, an iterative algorithm might be feasible.

As future work, we plan to apply our techniques to obtain
approximation algorithms for other behavioural pseudometrics such as,
for example, the one for systems that combine nondeterminism and
probability presented in \cite{DCPP05:qapl} and the pseudometric for
weak probabilistic bisimilarity in \cite{DGJP02:lics}.  In the latter
case the pseudometric can be characterized as the fixed point of a
functional based on the Kantorovich and Hausdorff metrics.  These can
easily be encoded in the first-order theory of the reals.  However, the
need to consider the transitive closure of the silent transition relation
suggests that some non-trivial extension of the work presented here is
called for.

\section*{Acknowledgement}

The authors would like to thank Christel Baier for providing
some pointers to the literature and Jeff Edmonds, Parke Godfrey
and the referees for their constructive feedback.

\bibliographystyle{plain}
\bibliography{submission}

\appendix

\section{Closure ordinal of $\Delta$}
\label{section:added}

The greatest fixed point of an order-preserving function on a complete lattice
can be obtained by iteration (see, for example, \cite[Exercise~4.13]{DP90}).

\begin{defi}
For each ordinal $\alpha$, the 1-bounded pseudometric $d^{\alpha}$ on $S$
is defined by
\begin{displaymath}
\begin{array}{rcll}
d^0 & = & \top\\
d^{\alpha + 1} & = & \Delta(d^{\alpha})\\
d^{\beta} & = & {\displaystyle \bigsqcap_{\alpha \in \beta} d^{\alpha}} & \mbox{if $\beta$ is a limit ordinal}
\end{array}
\end{displaymath}
\end{defi}

As we will see in the next example, for some systems we need at least $\omega$ iterations
to reach the greatest fixed point of $\Delta$.  

\begin{exa}
Consider the system of Example~\ref{example:1}.  For all $n$,
\begin{displaymath}
\begin{array}{rcl}
d_{n+1}(s_1, s_2) & = & \frac{1}{4} + \frac{5}{8} d_n(s_1, s_3)\\
d_{n+1}(s_1, s_3) & = & \frac{2}{5} d_n(s_2, s_3)\\
d_{n+1}(s_2, s_3) & = & \frac{1}{5} + \frac{7}{10} d_n(s_1, s_3)
\end{array}
\end{displaymath}
Hence, for this system we need $\omega$ iterations.
\end{exa}

In the rest of this appendix, we prove that we need at most
$\omega$ iterations for any system.  This tells us that the 
closure ordinal of $\Delta$ is
$\omega$, that is, $\Delta(d^{\omega}) = d^{\omega}$.
As a consequence, $d^{\omega}$ is the greatest fixed point of
$\Delta$ (see, for example, \cite[Example~4.13]{DP90}).  As we will 
see below, the fact that $d^{\omega}$ is a fixed point of $\Delta$ follows
from the facts that $\Delta$ is order-preserving 
(Proposition~\ref{proposition:5}) and Lipschitz 
(Proposition~\ref{proposition:d}).

In \cite[page~418]{DGJP02:lics}, Desharnais et al.\ state that
a functional similar to $\Delta$ has closure ordinal $\omega$.

Recall that for a pseudometric $d$, the equivalence relation $\equiv_d$ 
relates states that have distance zero.  From each equivalence class 
$[s]_d$ we pick a designated state which we denote by $<s>_d$.  
Hence, $<s>_d \in [s]_d$ and also $d(s, <s>_d) = 0$.

\begin{prop}
\label{proposition:a}
For all $s_1$, $s_2 \in S$,
\begin{displaymath}
d(<s_1>_d, <s_2>_d)
=
d(s_1, s_2).
\end{displaymath}
\end{prop}
\begin{proof}
\begin{eqnarray*}
\lefteqn{d(<s_1>_d, <s_2>_d)}\\
& \leq & d(<s_1>_d, s_1) + d(s_1, s_2) + d(s_2, <s_2>_d)\\
& = & d(s_1, s_2)\\
& \leq & d(s_1, <s_1>_d) + d(<s_1>_d, <s_2>_d) + d(<s_2>_d, s_2)\\
& = & d(<s_1>_d, <s_2>_d).
\end{eqnarray*}
\end{proof}

Let $d_1 \sqsubseteq d_2$.  The ratio $\rho(d_1, d_2)$
of $d_1$ and $d_2$ is defined by
\begin{displaymath}
\rho(d_1, d_2)
=
\min \left \{\, \frac{d_2(s_1, s_2)}{d_1(s_1, s_2)} \bigmid d_2(s_1, s_2) \gr 0  \, \right \}
\end{displaymath}
Note that we never divide by zero since $d_1 \sqsubseteq d_2$
and, hence, $d_1(s_1, s_2) \geq d_2(s_1, s_2)$.  

Below, we will use the convention that the minimum of 
the empty set is one and the maximum of the empty set is zero.

Given pseudometrics $d_1$ and $d_2$ such that $d_1 \sqsubseteq d_2$ and
given an $f \in (S, d_1) \ndi [0, 1]$, we next show that there
exists a $g_f \in (S, d_2) \to [0, 1]$ that is nonexpansive.

\begin{prop}
\label{proposition:b}
Let $d_1 \sqsubseteq d_2$ and $f \in (S, d_1) \ndi [0, 1]$.
Let $g_f : S \to [0, 1]$ be defined by
\begin{displaymath}
g_f(s) = \rho(d_1, d_2) f(<s>_{d_2}).
\end{displaymath}
Then $g_f \in (S, d_2) \ndi [0, 1]$.
\end{prop}
\begin{proof}
Let $s_1$, $s_2 \in S$.  We have to show that
\begin{displaymath}
| g_f(s_1) - g_f(s_2) | \leq d_2(s_1, s_2).
\end{displaymath}
We distinguish two cases.  If $d_2(s_1, s_2) = 0$ then
$<s_1>_{d_2} = <s_2>_{d_2}$ and, hence, $f(<s_1>_{d_2}) = f(<s_2>_{d_2})$.
Therefore $g_f(s_1) = g_f(s_2)$ and, hence, the property is vacuously
true.  Let $d_2(s_1, s_2) \gr 0$.  According to 
Proposition~\ref{proposition:a}, $d_2(<s_1>_{d_2}, <s_2>_{d_2}) \gr 0$.
Also $d_1(s_1, s_2) \gr 0$ since $d_1 \sqsubseteq d_2$, and
\begin{eqnarray*}
\lefteqn{| g_f(s_1) - g_f(s_2) |}\\
& = & | \rho(d_1, d_2) f(<s_1>_{d_2}) - \rho(d_1, d_2) f(<s_2>_{d_2}) |\\
& = & \rho(d_1, d_2) | f(<s_1>_{d_2}) - f(<s_2>_{d_2}) |\\
& \leq &  \rho(d_1, d_2) d_1(<s_1>_{d_2}, <s_2>_{d_2})
\comment{$f \in (S, d_1) \ndi [0, 1]$}\\
& \leq & \frac{d_2(<s_1>_{d_2}, <s_2>_{d_2})}{d_1(<s_1>_{d_2}, <s_2>_{d_2})} d_1(<s_1>_{d_2}, <s_2>_{d_2})\\
& = & d_2(<s_1>_{d_2}, <s_2>_{d_2})\\
& = & d_2(s_1, s_2)
\comment{Proposition~\ref{proposition:a}}
\end{eqnarray*}
\end{proof}

Next, we bound $f - g_f$ from above.

\begin{prop}
\label{proposition:c}
Let $d_1 \sqsubseteq d_2$ and $f \in (S, d_1) \ndi [0, 1]$.
Let $\mu = \min \{\, d_1(s_1, s_2) \mid d_1(s_1, s_2) \gr 0 \,\}$.
Then
\begin{displaymath}
f(s) - g_f(s)
\leq
\frac{\mu + 1}{\mu} \max_{s_1', s_2' \in S} d_1(s_1', s_2') - d_2(s_1', s_2')
\end{displaymath}
for all $s \in S$.
\end{prop}
\begin{proof}
Let $s \in S$. Then
\begin{eqnarray*}
\lefteqn{f(s) - g_f(s)}\\
& = & f(s) - \rho(d_1, d_2) f(<s>_{d_2})\\
& = & (f(s) - f(<s>_{d_2})) + (f(<s>_{d_2}) - \rho(d_1, d_2) f(<s>_{d_2})).
\end{eqnarray*}
Furthermore,
\begin{eqnarray*}
\lefteqn{f(s) - f(<s>_{d_2})}\\
& \leq & d_1(s, <s>_{d_2})
\comment{$f \in (S, d_1) \ndi [0, 1]$}\\
& = & d_1(s, <s>_{d_2}) - d_2(s, <s>_{d_2})
\comment{$d_2(s, <s>_{d_2}) = 0$}\\
& \leq & \max_{s_1', s_2' \in S} d_1(s_1', s_2') - d_2(s_1', s_2')
\end{eqnarray*}
and
\begin{eqnarray*}
\lefteqn{f(<s>_{d_2}) - \rho(d_1, d_2) f(<s>_{d_2})}\\
& \leq & 1 - \rho(d_1, d_2)\\
& = & 1 - \min \left \{\, \frac{d_2(s_1, s_2)}{d_1(s_1, s_2)} \bigmid d_2(s_1, s_2) \gr 0  \, \right \}\\
& = & \max \left \{\, \frac{d_1(s_1, s_2) - d_2(s_1, s_2)}{d_1(s_1, s_2)} \bigmid d_2(s_1, s_2) \gr 0  \, \right \}\\
& \leq & \frac{1}{\mu} \max \left \{\, d_1(s_1, s_2) - d_2(s_1, s_2) \mid d_2(s_1, s_2) \gr 0  \, \right \}\\
& \leq & \frac{1}{\mu} \max_{s_1', s_2' \in S} d_1(s_1', s_2') - d_2(s_1', s_2').
\end{eqnarray*}
\end{proof}

Now we can prove that $\Delta$ is Lipschitz, that is,
\begin{displaymath}
\max_{s_1, s_2 \in S} \Delta(d_1)(s_1, s_2) - \Delta(d_2)(s_1, s_2) 
\leq
\lambda \max_{s_1', s_2' \in S} d_1(s_1', s_2') - d_2(s_1', s_2').
\end{displaymath}
for some constant $\lambda$.

\begin{prop}
\label{proposition:d}
Let $d_1 \sqsubseteq d_2$.
For all $s_1$, $s_2 \in S$,
\begin{displaymath}
\Delta(d_1)(s_1, s_2) - \Delta(d_2)(s_1, s_2) 
\leq
|S| \frac{\mu + 1}{\mu} \max_{s_1', s_2' \in S} d_1(s_1', s_2') - d_2(s_1', s_2').
\end{displaymath}
\end{prop}
\begin{proof}
Let $s_1$, $s_2 \in S$.  Then
\begin{eqnarray*}
\lefteqn{\Delta(d_1)(s_1, s_2) - \Delta(d_2)(s_1, s_2)}\\
& = & \max\, \left \{\, \sum_{s \in S} f(s) (\pi(s_1, s) - \pi(s_2, s)) \bigmid f \in (S, d_1) \ndi [0, 1] \, \right \}
      -\\
&   & \max\, \left \{\, \sum_{s \in S} g(s) (\pi(s_1, s) - \pi(s_2, s)) \bigmid g \in (S, d_2) \ndi [0, 1] \, \right \}\\
& = & \max\, \left \{\, \min\, \left \{\, \sum_{s \in S} f(s) (\pi(s_1, s) - \pi(s_2, s)) - \sum_{s \in S} g(s) (\pi(s_1, s) - \pi(s_2, s)) \right . \right .\\
&   & \phantom{\max\, \left \{\, \min\, \left \{\, \right. \right .} \bigmid g \in (S, d_2) \ndi [0, 1] \,\Bigg\} \bigmid f \in (S, d_1) \ndi [0, 1] \,\Bigg\}\\
& = & \max\, \left \{\, \min\, \left \{\, \sum_{s \in S} (f(s) - g(s)) (\pi(s_1, s) - \pi(s_2, s)) \right . \right . \\
&   & \phantom{\max\, \left \{\, \min\, \left \{\, \right. \right .} \bigmid g \in (S, d_2) \ndi [0, 1] \,\Bigg\} \bigmid f \in (S, d_1) \ndi [0, 1] \,\Bigg\}\\
& \leq & \max\, \left \{\, \sum_{s \in S} (f(s) - g_f(s)) (\pi(s_1, s) - \pi(s_2, s)) \bigmid f \in (S, d_1) \ndi [0, 1] \, \right \}\\
&& \comment{Proposition~\ref{proposition:b}}\\
& \leq & \max\, \left \{\, \sum_{s \in S} f(s) - g_f(s) \bigmid f \in (S, d_1) \ndi [0, 1] \, \right \}\\
& \leq & |S| \frac{\mu + 1}{\mu} \max_{s_1', s_2' \in S} d_1(s_1', s_2') - d_2(s_1', s_2')
\comment{Proposition~\ref{proposition:c}}
\end{eqnarray*}
\end{proof}

Finally, we prove that the closure ordinal of $\Delta$ is $\omega$.

\begin{prop}
\label{proposition:6a}
$\Delta(d^{\omega}) = d^{\omega}$.
\end{prop}
\begin{proof}
First, we show that $\Delta(d^{\omega}) \sqsubseteq d^{\omega}$.
By definition, $d^{\omega} = \bigsqcap_{n \in \omega} d^n \sqsubseteq d^n$ for
all $n \in \omega$.  Since $\Delta$ is order-preserving,
$\Delta(d^{\omega}) \sqsubseteq \Delta(d^n) = d^{n+1}$ for all $n \in \omega$.
Obviously, $\Delta(d^{\omega}) \sqsubseteq d^0$.  Therefore,
$\Delta(d^{\omega})$ is a lower bound of $\{\, d^n \mid n \in \omega \,\}$.
Since $d^{\omega}$ is the greatest lower bound by definition,
$\Delta(d^{\omega}) \sqsubseteq d^{\omega}$.

We have left to show that $\Delta(d^{\omega}) \sqsupseteq d^{\omega}$, that is,
$\Delta(d^{\omega})(s_1, s_2) \leq d^{\omega}(s_1, s_2)$
for all $s_1$, $s_2 \in S$.  Let $s_1$, $s_2 \in S$.  Let $\epsilon \gr 0$.  It 
suffices to show that there exists an $n$ such that
$\Delta(d^{\omega})(s_1, s_2) - d^{n+1}(s_1, s_2) \leq \epsilon$. 
Let $\mu = \min \{\, d^{\omega}(s_1, s_2) \mid d^{\omega}(s_1, s_2) \gr 0 \,\}$.
Since the set $S$ is finite, for every $\delta \gr 0$ there exists an $n$
such that for all $s_1'$, $s_2' \in S$,
\begin{displaymath}
d^{\omega}(s_1', s_2') - d^n(s_1', s_2') \leq \delta.
\end{displaymath}
Here we pick $\delta$ to be $\frac{\mu \epsilon}{(\mu + 1) |S|}$.
From Proposition~\ref{proposition:d} we can conclude that
\begin{eqnarray*}
\lefteqn{\Delta(d^{\omega})(s_1, s_2) - d^{n+1}(s_1, s_2)}\\
& = & \Delta(d^{\omega})(s_1, s_2) - \Delta(d^n)(s_1, s_2)\\
& \leq & \epsilon.
\end{eqnarray*}
\end{proof}

\section{An implementation in Mathematica}
\label{section:8}

A decision procedure for the first order theory of real closed fields based 
on quantifier elimination was first given by Tarski \cite{T51}.  A number of 
algorithms have been developed thereafter for the theory (see, for example,
\cite{SPR96:jacm,C75:atfl,H05}).  Collin's algorithm is implemented in 
the tool Mathematica and can be used for solving our formulae.  However, it 
works for very small examples and therefore it is essential to simplify the 
formula and reduce its size to make it solvable.  To simplify the formula,
we first compute some of the distances using the following results.

\begin{prop}
\label{proposition:A}
\mbox{}
\begin{enumerate}[$\bullet$]
\item
If $s_1 \not\rightarrow$ and $s_2 \not\rightarrow$ then $d_1(s_1, s_2) = 0$.
\item
If $s_1 \not\rightarrow$ and $s_2 \rightarrow$, or $s_1 \rightarrow$ and 
$s_2 \not\rightarrow$ then $d_1(s_1, s_2) = 1$.
\end{enumerate}
\end{prop}
\begin{proof}
We only consider the first case.  The second one can be proved similarly.
If $s_1 \not\rightarrow$ and $s_2 \not\rightarrow$ then
$\delta(s_1, s_2) = \Delta(\delta)(s_1, s_2) = 0$.
\end{proof}

\begin{exa}
Consider the probabilistic transition system of Example~\ref{example:1}.
State $s_4$ has distance one to all other states.
\end{exa}

Next, we present a simple characterization of the distance between a state 
that never terminates (that is, the probability of reaching a state with no 
outgoing transitions is zero) and another state.

Given a state $s$ and $n \in \omega + 1$, $\tau_n(s)$ is the probability
of terminating in less than $n$ transitions when started in $s$.

\begin{defi}
\label{definition:termination}
For each $n \in \omega + 1$, the function $\tau_n : S \to [0, 1]$
is defined by
\begin{displaymath}
\begin{array}{rcl}
\tau_0(s) & = & 0\\
\tau_{n+1}(s) & = & \left \{ 
\begin{array}{ll} 
1 & \mbox{if $s \not\rightarrow$}\\
\sum_{s' \in S} \pi(s, s') \tau_n(s') & \mbox{otherwise}
\end{array} \right . \\ 
\tau_{\omega}(s) & = & \sup_{n \in \omega} \tau_n(s)
\end{array}
\end{displaymath}
\end{defi}

\begin{exa}
Consider the probabilistic transition system of Example~\ref{example:1}.
Then we have that $\tau_{\omega}(s_1) = \frac{1}{9}$, 
$\tau_{\omega}(s_2) = \frac{5}{18}$, $\tau_{\omega}(s_3) = 0$,
$\tau_{\omega}(s_4) = 1$ and $\tau_{\omega}(s_5) = 0$.
\end{exa}

Obviously, for a state $s$ without outgoing transitions, we have
that $\tau_{\omega}(s) = 1$.  For a state $s$ that cannot reach
any state without outgoing transitions, we have that $\tau_{\omega}(s) = 0$.
For the remaining states, we can compute the probability of termination
using standard techniques as described in, for example, 
\cite[Section~11.2]{GS97}.

\begin{prop}
\label{proposition:7a}
If $\tau_{\omega}(s_2) = 0$ then $d_1(s_1, s_2) = \tau_{\omega}(s_1)$.
\end{prop}
\proof%\begin{proof}
Assume that $\tau_{\omega}(s_2) = 0$.  We prove that for all $n \in \omega + 1$,
\begin{displaymath}
d^n(s_1, s_2) = \tau_n(s_1)
\end{displaymath}
by induction on $n$.
\begin{enumerate}[$\bullet$]
\item
Obviously, $d^0(s_1, s_2) = 0 = \tau_0(s_1)$.
\item
We have to prove that $d^{n+1}(s_1, s_2) = \tau_{n+1}(s_1)$.  We
distinguish the following two cases.
\begin{enumerate}[$-$]
\item
If $s_1 \not\rightarrow$
then $d^{n+1}(s_1, s_2) = 1 = \tau_{n+1}(s_1)$.
\item
Now let us assume that $s_1 \rightarrow$.  First we show that
$\tau_n$ as a function from $(S, d^n)$ to $[0, 1]$ is nonexpansive.
For all $s$, $s'$,
\begin{eqnarray*}
|\tau_n(s) - \tau_n(s')|
& = & | d^n(s, s_2) - d^n(s', s_2) |
\comment{induction}\\
& \leq & d^n(s, s')
\comment{triangle inequality}
\end{eqnarray*}
Since
\begin{eqnarray*}
\lefteqn{d^{n+1}(s_1, s_2)}\\
& = & \Delta(d^n)(s_1, s_2)\\
& \geq & \sum_{s \in S} \tau_n(s) (\pi(s_1, s) - \pi(s_2, s))
\comment{$\tau_n$ is nonexpansive}\\
& = & \sum_{s \in S} \tau_n(s) \pi(s_1, s) - \sum_{s \in S} \tau_n(s) \pi(s_2, s)\\
& = & \tau_{n+1}(s_1) - \tau_{n+1}(s_2)\\
& = & \tau_{n+1}(s_1)
\comment{$\tau_{\omega}(s_2) = 0$ and, hence, $\tau_{n+1}(s_2) = 0$}
\end{eqnarray*}

Let $f \in (S, d^n) \ndi [0, 1]$.  For all $s$,
\begin{displaymath}
f(s) - f(s_2) \leq | f(s) - f(s_2) | \leq d^n(s, s_2) = \tau_n(s).
\end{displaymath}
As a consequence,
\begin{eqnarray*}
\lefteqn{\sum_{s \in S} f(s) (\pi(s_1, s) - \pi(s_2, s))}\\
& = & \sum_{s \in S} f(s) \pi(s_1, s) - \sum_{s \in S} f(s) \pi(s_2, s)\\
& = & \sum_{s \in S} (f(s) - f(s_2)) \pi(s_1, s) - \sum_{s \in S} (f(s) - f(s_2)) \pi(s_2, s)\\
&& \comment{$\sum_{s \in S} f(s_2) \pi(s_i, s) = f(s_2)$}\\
& = & \sum_{s \in S} (f(s) - f(s_2)) (\pi(s_1, s) - \pi(s_2, s))\\
& \leq & \sum_{s \in S} \tau_n(s)(\pi(s_1, s) - \pi(s_2, s))\\
& = & \tau_{n+1}(s_1).
\end{eqnarray*}
Since $f$ was chosen arbitrarily, we can conclude that
\begin{displaymath}
d^{n+1}(s_1, s_2) \leq \tau_{n+1}(s_1).
\end{displaymath}
\item
Finally,
\begin{eqnarray*}
d^{\omega}(s_1, s_2)
& = & \sup_n d^n(s_1, s_2)\\
& = & \sup_n \tau_n(s_1) 
\comment{by induction}\\
& = & \tau_{\omega}(s_1).
\end{eqnarray*}
\end{enumerate}
From Theorem~\ref{proposition:6} and Proposition~\ref{proposition:6a} we can conclude 
that $d_1(s_1, s_2) = d^{\omega}(s_1, s_2) = \tau_{\omega}(s_1)$.\qed
\end{enumerate}

\begin{exa}
Consider the probabilistic transition system of Example~\ref{example:1}.
From Proposition~\ref{proposition:7a} we can conclude that
$d_1(s_1, s_3) = \frac{1}{9}$, $d_1(s_2, s_3) = \frac{5}{18}$,
$d_1(s_4, s_3) = 1$ and $d_1(s_5, s_3) = 0$.
\end{exa}

Given a probabilistic bisimulation $\mathcal{R}$, we can quotient the probabilistic
transition system $<S, \pi>$ as follows.

\begin{defi}
Let $\mathcal{R}$ be a probabilistic bisimulation.  The probabilistic transition
system $<S_{\mathcal{R}}, \pi_{\mathcal{R}}>$ consists of
\begin{enumerate}[$\bullet$]
\item
the set $S_{\mathcal{R}} = \{\, [s] \mid s \in S \,\}$ of $\mathcal{R}$-equivalence classes and 
\item
the function $\pi_{\mathcal{R}} : S_{\mathcal{R}} \times S_{\mathcal{R}} \to [0, 1]$ defined by
\begin{displaymath}
\pi_{\mathcal{R}}([s], [s']) = \sum_{s'' \mathbin{\mathcal{R}} s'} \pi(s, s'').
\end{displaymath}
\end{enumerate}
\end{defi}

Note that the function $\pi_{\mathcal{R}}$ is well-defined since $\mathcal{R}$ 
is a probabilistic bisimulation.  We will apply the above quotient 
construction for probabilistic bisimilarity (which can be computed in
polynomial time \cite{BEM00:jcss}).

\begin{exa}
Consider the probabilistic transition system of Example~\ref{example:1}.
%According to Proposition~\ref{proposition:8}, 
The smallest equivalence relation containing $\{ <s_3, s_5> \}$ is a 
probabilistic bisimulation.  The resulting quotient can be depicted as
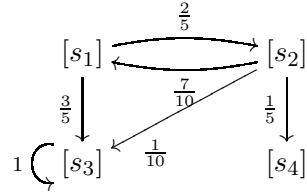
\begin{displaymath}
\UseComputerModernTips
\xymatrix@C=5em{
[s_1] \ar@/^/[r]^{\frac{2}{5}} \ar[d]_{\frac{3}{5}} & [s_2] \ar@/^/[l]^{\frac{7}{10}} \ar[d]_{\frac{1}{5}} \ar[dl]^(0.7){\frac{1}{10}}\\
[s_3] \ar@(ul,dl)_{1} & [s_4]}
\end{displaymath}
\end{exa}

By quotienting, the number of states that need to be considered and, hence,
the number of variables in the formula may be reduced.  However, we still
have to check that the quotiented system gives rise to the same distances.
Next we relate the behavioural pseudometric $d_1$ of the original system 
$<S, \pi>$ with the behavioural pseudometric $d_{\mathcal{R}}$ of the 
quotiented system $<S_{\mathcal{R}}, \pi_{\mathcal{R}}>$.

\begin{prop}
\label{proposition:B}
For all $s_1$, $s_2 \in S$, $d_{\mathcal{R}}([s_1], [s_2]) = d_1(s_1, s_2)$.
\end{prop}
\proof%\begin{proof}
First all, note that
\begin{displaymath}
\sum_{s' \in S} \pi(s, s')
= \sum_{[s'] \in S_{\mathcal{R}}} \sum_{s'' \mathbin{\mathcal{R}} s'} \pi(s, s'')\\
= \sum_{[s'] \in S_{\mathcal{R}}} \pi_{\mathcal{R}}([s], [s']).
\end{displaymath}
As a consequence, we have left to consider the case $s_1 \rightarrow$
and $s_2 \rightarrow$.  We prove that for all $n \in \omega + 1$,
$d_{\mathcal{R}}^n([s_1], [s_2]) = d_1^n(s_1, s_2)$ by induction on $n$.
We distinguish the following three cases.
\begin{enumerate}[$\bullet$]
\item
If $n = 0$ then the property is vacuously true.
\item
Assume that $d_{\mathcal{R}}^n([s_1'], [s_2']) = d_1^n(s_1', s_2')$ for all $s_1'$, $s_2' \in S$.
Let $s_1$, $s_2 \in S$.  We have to prove that $d_{\mathcal{R}}^{n+1}([s_1], [s_2]) = d_1^{n+1}(s_1, s_2)$.
In the proof of this case, we make use of the following two observations.  
For each $f \in (S_{\mathcal{R}}, d_{\mathcal{R}}^n) \ndi [0, 1]$, there exists a $g \in (S, d_1^n) \ndi [0, 1]$
such that $g(s) = f([s])$ for all $s \in S$, since
\begin{eqnarray*}
| g(s) - g(s') |
& = & | f([s]) - f([s']) |\\
& \leq & d_{\mathcal{R}}^n(s, s') 
\comment{$f$ is nonexpansive}\\
& = & d_1^n(s, s') 
\comment{induction}.
\end{eqnarray*} 
Similarly, we can show that for each $g \in (S, d_1^n) \ndi [0, 1]$, there exists
$f \in (S_{\mathcal{R}}, d_{\mathcal{R}}^n) \ndi [0, 1]$ such that $f([s]) = g(s)$ for all $s \in S$.
Note that if states $s$ and $s'$ are probabilistic bisimilar then 
$d_1(s, s') = 0$ and, hence, $d_1^n(s, s') = 0$ and, therefore, $g(s) = g(s')$,
since $g$ is nonexpansive.
\begin{eqnarray*}
\lefteqn{d_{\mathcal{R}}^{n+1}([s_1], [s_2])}\\
& = & \Delta(d_{\mathcal{R}}^n)([s_1], [s_2])\\
& = & \max\, \left \{\, \sum_{[s] \in S_{\mathcal{R}}} f([s]) (\pi_{\mathcal{R}}([s_1], [s]) - \pi_{\mathcal{R}}([s_2], [s])) \bigmid f \in (S_{\mathcal{R}}, d_{\mathcal{R}}^n) \ndi [0, 1] \, \right \}\\
& = & \max\, \left \{\, \sum_{[s] \in S_{\mathcal{R}}} f([s]) \sum_{s' \mathbin{\mathcal{R}} s} (\pi(s_1, s') - \pi(s_2, s')) \bigmid f \in (S_{\mathcal{R}}, d_{\mathcal{R}}^n) \ndi [0, 1] \, \right \}\\
& = & \max\, \left \{\, \sum_{[s] \in S_{\mathcal{R}}} \sum_{s' \mathbin{\mathcal{R}} s} f([s']) (\pi(s_1, s') - \pi(s_2, s')) \bigmid f \in (S_{\mathcal{R}}, d_{\mathcal{R}}^n) \ndi [0, 1] \, \right \}\\
& = & \max\, \left \{\, \sum_{s \in S} g(s) (\pi(s_1, s) - \pi(s_2, s)) \bigmid g \in (S, d_1^n) \ndi [0, 1] \, \right \}\\
& = & \Delta(d_1^n)(s_1, s_2)\\
& = & d_1^{n+1}(s_1, s_2).
\end{eqnarray*}
\item
Furthermore,
\begin{eqnarray*}
d_{\mathcal{R}}^{\omega}([s_1], [s_2])
& = & \sup_n d_{\mathcal{R}}^n([s_1], [s_2])\\
& = & \sup_n d_1^n(s_1, s_2)
\comment{induction}\\
& = & d_1^{\omega}(s_1, s_2).\rlap{\hbox to186 pt{\hfill\qEd}}%\eqno{\qEd}
\end{eqnarray*}
\end{enumerate}

To simplify the formula even further, we exploit the following three observations.
\begin{enumerate}[$\bullet$]
\item
Since $d$ is a pseudometric, $d(s_i, s_i) = 0$ and 
$d(s_i, s_j) = d(s_j, s_i)$.  Therefore,
in $\mbox{\rm pseudo}(d) \wedge \mbox{\rm post-fixed}(d)$ we can replace
all $d_{ii}$'s with zero and all $d_{ij}$'s where $i \gr j$ with
$d_{ji}$'s.  As a consequence, we only need to consider $d_{ij}$'s with $i \ls j$.
This reduces the number of variables in the formula considerably.
\item
Let $C$ be the set of pairs of states for which the distances have already
been computed.  Then 
\begin{displaymath}
\exists d\, \mbox{\rm pseudo}(d) \wedge \mbox{\rm post-fixed}(d) \wedge d_{i_0 j_0} \leq m
\end{displaymath}
is equivalent to
\begin{displaymath}
\exists d\, \mbox{\rm pseudo}(d) \wedge \mbox{\rm post-fixed}(d) \wedge d_{i_0 j_0} \leq m \wedge \bigwedge_{(i, j) \in C} d_{ij} = d_1(s_i, s_j)
\end{displaymath}
since $d_1$ is the greatest post-fixed point.  As a consequence, we can
replace all $d_{ij}$'s where $(i, j) \in C$ with their already computed
distances $d_1(s_i, s_j)$.  Again, the number of variables may be reduced.
\item
If $\pi_{i_0 j} = 0$, we can infer that $\mu_{ij} = 0$ for all $1 \leq i \leq N$. 
As a consequence, we can replace the occurrences of all those $\mu_{ij}$'s 
with 0.  Symmetrically, if $\pi_{j_0 i} = 0$ we can simplify the formula
similarly.  Also this simplification may reduce the number of variables.
\end{enumerate}

We have implemented these simplifications in the form of a Java program that takes
as input the probability matrix $\pi$ and that produces as output the simplified
formula in a format that can be fed to Mathematica.%
\footnote{The code and documentation is available at the URL\ 
{\tt www.cse.yorku.ca/$\raisebox{-0.5ex}{\mbox{\tt \~{}}}$franck/research/pm2m}.}

\begin{exa}
Consider the probabilistic transition system of Example~\ref{example:1}.
The simplified formula for this system is given below.
\lstset{basicstyle=\ttfamily,basewidth={0.5em,0.1em},breaklines=true,mathescape=true}

\lstset{basicstyle=\ttfamily,
        breaklines=true,
        basewidth={0.5em,0.1em},
        numbers=left, 
        numberstyle=\tiny}
\begin{small}
\begin{lstlisting}
Reduce[
  Exists[d12,
    (0 <= d12 <= 1) && (0.11112 <= d12 + 0.27778) && (d12 <= 0.38889) && 
    Exists[{u12,u13,u32,u42,u43,u33}, 
      (0 <= u12 <= 1) && (0 <= u13 <= 1) && (0 <= u32 <= 1) && 
      (0 <= u42 <= 1) && (0 <= u43 <= 1) && 
      (u12 + u32 + u42 == 0.4) && (u13 + u43 + u33 == 0.6) && 
      (u12 + u13 == 0.7) && (u32 + u33 == 0.1) && (u42 + u43 == 0.2) && 
      (d12 * u12 + 0.11112 * u13 + 0.27778 * u32 + u42 + u43 <= d12)] && 
    Exists[{u21,u23,u24,u31,u33, u34}, 
      (0 <= u21 <= 1) && (0 <= u23 <= 1) && (0 <= u24 <= 1) && 
      (0 <= u31 <= 1) && (0 <= u34 <= 1) && 
      (u21 + u31 == 0.7) && (u23 + u33 == 0.1) && (u24 + u34 == 0.2) && 
      (u21 + u23 + u24 == 0.4) && (u31 + u33 + u34 == 0.6) && 
      (d12 * u21 + 0.27778 * u23 + u24 + 0.11112 * u31 + u34 <= d12)] && 
    (0 <= d12 <= 0.5)]]
\end{lstlisting}
\end{small}
Line~3 correspond to $\mbox{pseudo}(d)$, line 4--9 correspond to
$\mbox{post-fixed}_1(d, 1, 2)$ and line 10--15 correspond to
$\mbox{post-fixed}_1(d, 2, 1)$.  
The formula was reduced to true by Mathematica in 8.2 seconds on a 3GHz machine with 1GB RAM.
When feeding Mathematica the formula that has not been simplified, it runs out of
memory after some time.

We also attempted to solve this example with a solver called QEPCAD~B \cite{C03:jsac} but
the performance of Mathematica on this example was better.
\end{exa}

\end{document}